\documentclass[pdflatex,sn-mathphys-num]{sn-jnl}

\usepackage{graphicx}%
\usepackage{multirow}%
\usepackage{amsmath,amssymb,amsfonts}%
\usepackage{amsthm}%
\usepackage{mathrsfs}%
\usepackage[title]{appendix}%
\usepackage{xcolor}%
\usepackage{textcomp}%
\usepackage{manyfoot}%
\usepackage{booktabs}%
\usepackage{algorithm}%
\usepackage{algorithmicx}%
\usepackage{algpseudocode}%
\usepackage{listings}%
\usepackage{epstopdf}%
\usepackage{mathrsfs}

\usepackage{graphicx} 
\usepackage{subcaption} 

\theoremstyle{thmstyleone}%
\theoremstyle{thmstyletwo}%

\theoremstyle{thmstylethree}%

\begin{document}

\title[Article Title]{Magnon-squeezing–induced nonreciprocal quantum coherence in a cavity magnomechanical system 
}


\author*[1]{\fnm{Abdelkader} \sur{Hidki}}\email{abdelkader.hidki@gmail.com} 
\author[2]{\fnm{Amjad} \sur{Sohail}}
\author[3]{\fnm {Tesfay Gebremariam}\sur{Tesfahannes}}
\author[4]{\fnm{Mulugeta Tadesse}\sur{Bedore}}  
\author[1]{\fnm {Mostafa}\sur{Nassik}} 

\affil[1]{\orgdiv{LPTHE, Department of Physics, Faculty of Sciences}, \orgname{Ibn Zohr University},  \city{Agadir}, \postcode{80000}, \country{Morocco}}

\affil[2]{\orgdiv{Department of Physics}, \orgname{Government College University, Allama Iqbal Road},  \city{Faisalabad}, \postcode{38000}, \country{Pakistan}}

\affil[3]{\orgdiv{Department of Physics}, \orgname{Arba Minch University},  \city{Arba Minch}, \postcode{21}, \country{Ethiopia}}
\affil[4]{\orgdiv{Department of Physics}, \orgname{Wachemo University}, \city{Hossana}, \postcode{667}, \country{Ethiopia}}


\abstract{We investigate quantum coherence in a hybrid cavity magnomechanical system incorporating a squeezed-magnon drive. By analyzing the Gaussian quantum coherence of the cavity, magnonic, and mechanical subsystems, as well as the total system coherence, we identify the critical roles of phase control, coupling strength, drive power, and thermal noise. We show that the squeezing amplitude and phase precisely modulate the effective magnon frequency and damping, enabling phase-dependent enhancement and nonreciprocal transfer of coherence. Our systematic parameter analysis indicates that increasing driving power and photon–magnon coupling enhances quantum coherence, while thermal decoherence leads to its degradation. However, this effect is partially suppressed by the presence of magnon squeezing. The results show that squeezed magnons are a robust and tunable resource for controlling, stabilizing, and optimizing quantum coherence in cavity magnomechanical platforms, offering potential applications in hybrid magnonic systems and coherent quantum information processing.}

\keywords{Nonreciprocal quantum coherence, Single-mode coherence, Cavity magnomechanics,  YIG sphere, Decoherence.}


\maketitle
\section{Introduction}
Recently, hybrid quantum systems that integrate distinct physical platforms with complementary properties have emerged as a key paradigm for advancing practical quantum information technologies \cite{xiang2013hybrid, xie2021triangle,ladd2010quantum}. Exploiting the unique features of each subsystem allows these hybrid systems to achieve functionalities unattainable in a single platform \cite{clerk2020hybrid}. Among these, the cavity magnomechanical system has emerged as a rapidly growing area of research. This system integrates a microwave cavity with a yttrium iron garnet (YIG) sphere, in which the collective spin excitations (magnons) interact strongly with the cavity’s microwave photons as a result of the YIG’s exceptionally high spin density and strong spin–spin exchange interaction \cite{lachance2019hybrid, lakhfif2024maximum}. In high-quality cavities, this interaction can even enter the ultrastrong coupling regime \cite{makihara2021ultrastrong,li2020phase,bedore2025enhancement}, leading to the formation of magnon-cavity polaritons through the collective excitation of a large ensemble of spins \cite{wang2020dissipative}. Furthermore, the magnon mode couples to the mechanical vibration mode of the YIG sphere through the magnetostrictive interaction, in which the spin dynamics induce mechanical deformation, thereby establishing an intrinsic coupling between the magnetic and mechanical degrees of freedom \cite{zhang2016cavity}. Meanwhile, compared with optomechanical systems \cite{aspelmeyer2014cavity}, cavity magnomechanical platforms offer higher tunability and lower dissipation, providing a promising platform for realizing quantum information processing and exploring macroscopic quantum phenomena in hybrid quantum systems \cite{li2020noise, wang2022entanglement, hidki2023enhanced, tadesse2024distant}.

 Cavity magnomechanical systems constitute a versatile platform for exploring a broad range of intriguing quantum phenomena. These include entanglement \cite{li2018magnon, osada2018brillouin, viola2016coupled, xu2020floquet, hidki2025entanglement}, asymmetric steering \cite{hidki2024asymmetric}, magnon squeezing \cite{zhang2021generation}, magnon-squeezing-enhanced slow light \cite{lu2023magnon,kong2019magnetically},  magnon lasing and chaos, high precision magnetometry and thermometry \cite{potts2020magnon}, magnon-squeezing-induced ground-state cooling of mechanical modes \cite{asjad2023magnon}, and quantum state storage and retrieval \cite{sarma2021cavity}, among others. Furthermore, in quantum precision measurement and quantum information processing, squeezed states, which are fundamental nonclassical resources, play a crucial role \cite{braunstein2005quantum}. For instance, squeezed light has been employed to enhance the sensitivity of interferometric gravitational-wave detectors \cite{zhao2020weak}, generate entangled resources for quantum teleportation \cite{furusawa1998unconditional}, and probe the quantum-classical boundary \cite{haroche2014controlling}. At the same time, magnonic squeezing can be realized through several approaches, such as exploiting the intrinsic nonlinearity of magnetostriction \cite{li2023squeezing}, employing reservoir-engineered cavity magnomechanical schemes \cite{qian2024strong}, or utilizing two-tone qubit driving techniques \cite{guo2023magnon}. Moreover, cavity magnomechanical systems provide a powerful platform for exploring a broad spectrum of nonlinear phenomena. Theoretically, they have been proposed for the realization of nonreciprocal entanglement \cite{kong2024nonreciprocal,imara2025squeezed}, entangled and squeezed states \cite{li2019squeezed, zou2020tuning}, steady Bell-state generation \cite{yuan2020steady}, stationary one-way quantum steering \cite{wei2024controlling},  and nonreciprocal high-order sidebands \cite{wang2021nonreciprocal}. More recently, nonreciprocal behavior in such systems has been achieved by using mechanisms that induce frequency shifts in magnon resonances \cite{ono2015barnett}.

 Quantum coherence, originating from the principle of quantum superposition, plays a central role in quantum information science, quantum thermodynamics, and quantum optics \cite{adesso2007entanglement, singh2021entanglement}. It is a fundamental property of quantum physics, providing profound insight into a wide range of quantum phenomena. It should be noted that several methods have been proposed to quantify quantum coherence, such as measures based on geometric distance or concurrence \cite{streltsov2015measuring} and relative entropy \cite{xu2016quantifying}. Quantifying coherence provides valuable insight into the boundary between quantum and classical domains and facilitates the study of how coherence governs correlations between quantum systems \cite{buchmann2015quantum}. In recent years, macroscopic quantum coherence has been extensively investigated in various physical systems, including optomechanical platforms \cite{zheng2016detecting,zhao2018coherence,li2019quantum,mekonnen2024enhancing,jin2021macroscopic} and Josephson junctions \cite{shalibo2010lifetime}.
Furthermore, magnomechanical systems incorporating a YIG sphere constitute an excellent platform for investigating quantum coherence at the macroscopic scale. For instance, quantum coherence and its nonreciprocity have been explored in systems consisting of a spinning microwave resonant cavity and a YIG sphere \cite{zhang2024nonreciprocal}; bipartite entanglement and macroscopic quantum coherence have been studied in a hybrid optomechanical Laguerre–Gaussian rotational cavity incorporating two YIG spheres \cite{hidki2024entanglement}; and macroscopic quantum coherence and tripartite entanglement have been examined in a three-mode magnomechanical system \cite{qiu2022controlling}. More recently, nonreciprocal quantum coherence based on the Barnett effect has also been investigated \cite{jia2025nonreciprocal}. Therefore, proposing a scheme that employs magnon squeezing to achieve nonreciprocal quantum coherence in a cavity magnomechanical system represents an avenue that remains unexplored in such systems.

To address this gap, we consider a cavity magnomechanical system to demonstrate how magnon squeezing induces nonreciprocal quantum coherence. Here, the magnon couples to a phonon via the magnetostrictive interaction and to a photon via the magnetic dipole interaction. We first quantify the Gaussian quantum coherence in each subsystem and the total system. We then show how this coherence depends on the phase, photon-magnon coupling, drive power, and thermal noise. Our key finding is that the squeezing amplitude and phase directly control the magnon frequency and damping. This control allows us to enhance and transfer nonreciprocal coherence between the three modes.\\
Furthermore, we find that higher drive power and stronger coupling enhance coherence, while thermal noise degrades it. However, magnon squeezing can reduce this thermal degradation. These results establish squeezed magnons as a robust, tunable resource for controlling, stabilizing, and optimizing quantum coherence in cavity magnomechanical systems, opening promising avenues for hybrid magnonic technologies and coherent quantum information processing.

The remainder of this paper is structured as follows. In Section \ref{sec2}, we present the physical model of the system and its dynamical equations. In Section \ref{sec3}, we quantify the single-mode quantum coherence as well as the total coherence. Section \ref{sec4} presents our main results and their analysis. Finally, Section \ref{sec5} provides a concise summary of our findings.

	\section{Physical model and dynamical of the system}\label{sec2}


	\begin{figure}[tbh!]
	\centerline{\includegraphics[width=13cm]{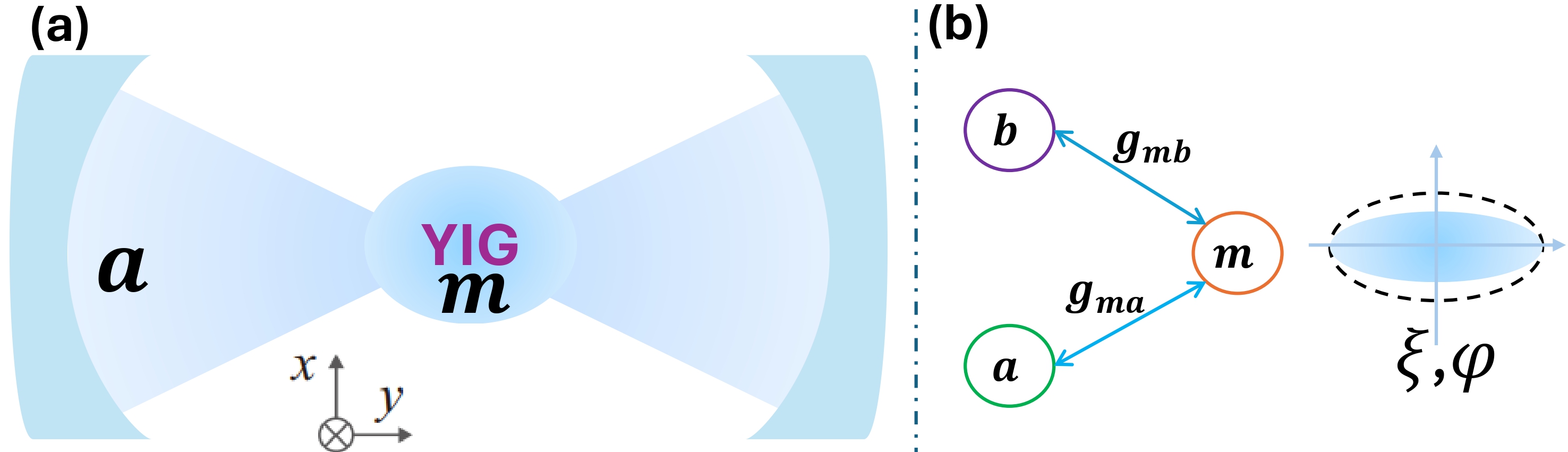}}
	\caption{\footnotesize{(a) Schematic diagram of the cavity magnomechanical system consisting of a microwave cavity coupled to a YIG sphere. The YIG sphere supports both magnon ($m$) and phonon ($b$) modes. It is positioned at the magnetic field maximum of the cavity mode ($a$) while being subjected to a uniform bias magnetic field along the $z$ axis, which enables photon–magnon coupling. The magnon–phonon interaction arises from the intrinsic magnetostrictive effect. (b) Equivalent mode coupling representation of the system, where $g_{ma}$ and $g_{mb}$ denote the photon–magnon and magnon–phonon coupling strengths, respectively. The light-blue ellipse illustrates the squeezed magnon mode characterized by the squeezing parameter $\xi$ and phase $\varphi$, which can be used to enhance nonreciprocal quantum coherence.
	}}
	\label{F1}
\end{figure}

As shown in Fig. \ref{F1}, we consider a cavity magnomechanical system consisting of a magnon mode, a microwave cavity mode, and a mechanical vibration mode. When the frequency of the magnon mode is tuned into resonance with that of the microwave photons, strong magnon–photon coupling occurs. Simultaneously, the magnon mode interacts with the mechanical vibration mode through the magnetostrictive coupling mechanism. Typically, the magnetostrictive interaction is governed by the resonance frequencies of the magnon and phonon modes \cite{zhang2016cavity}. In this case, since the magnon frequency is much higher than the mechanical frequency, the interaction between the magnon and phonon modes becomes dispersive. Moreover, because the YIG sphere is much smaller than the wavelength of the microwave fields, the effect of radiation pressure can be neglected \cite{li2018magnon}.
In the rotating frame with respect to the driving magnetic field frequency $\omega_l$, the Hamiltonian of the hybrid system can be written as:
\begin{eqnarray}\label{E1}
	\label{1}
		\mathcal{H}&=&\Delta_a a^\dagger a +\Delta_{m} m^\dagger m+\frac{\omega_{b}}{2}(q^2+p^2)  
        + g_{mb}m^\dagger mq 
		+ g_{ma} (a^\dagger m+  m^\dagger a) \nonumber\\
        &+&  i\Omega_l(m^\dagger-m ) + \frac{i}{2} \xi(m^{\dagger 2} e^{i\varphi} - m^{2} e^{-i\varphi}),   
\end{eqnarray}
where $a$ ($a^\dagger$) and $m$ ($m^\dagger$) denote the annihilation (creation) operators of the cavity and magnon modes, respectively. They obey the canonical bosonic commutation relations $[o, o^\dagger] = 1,  o = a, m$, with corresponding resonance frequencies $\omega_a$ and $\omega_m$. The parameters $\Delta_{a(m)} = \omega_{a(m)} - \omega_l$ represent the detuning between the mode frequencies $\omega_{a(m)}$ and the driving frequency $\omega_l$. In addition, the magnon frequency $\omega_m$ can be easily adjusted by varying the external bias magnetic field $H$ according to $\omega_m = \Gamma H$, where the gyromagnetic ratio is $\Gamma = 28,{\rm GHz/T}$ \cite{li2018magnon}. The operators $q$ and $p$ denote the dimensionless position and momentum, respectively, which satisfy the canonical commutation relation $[q, p] = i$ and are associated with the mechanical resonance frequency $\omega_b$. The coupling rates of the magnon–cavity and magnon–phonon interactions are denoted by $g_{ma}$ and $g_{mb}$, respectively. Furthermore, $\Omega_l = \sqrt{2P_l \gamma_m / \hbar \omega_l}$ characterizes the strength of the magnon driving field, determined by the input power $P_l$, the drive frequency $\omega_l$, and the magnon decay rate $\gamma_m$ \cite{jia2025nonreciprocal}. Moreover, the final term in the Hamiltonian of Eq.~\ref{E1} represents the magnon-squeezing interaction, characterized by the squeezing amplitude $\xi$ and phase $\varphi$. Such magnon squeezing can be achieved through several experimental techniques, including transferring squeezing from an optical cavity driven by applying two-tone microwave fields to excite the magnon mode \cite{guo2023magnon}, a squeezed vacuum field \cite{li2019squeezed,hidki2023transfer}, exploiting the intrinsic magnetic anisotropy of ferromagnetic materials \cite{kamra2016super}, or utilizing magnetostrictive nonlinearities \cite{asjad2023magnon,lu2023magnon}.

By applying the Heisenberg equations of motion and incorporating the corresponding damping and noise terms, the quantum Langevin equations (QLEs) of the system are expressed as follows:
\begin{eqnarray}
	\label{2}
	\dot{a} &= &-( \gamma_{a}+i\Delta_a ) a - i g_{ma} m + \sqrt{2\gamma_a} a^\text{in}, \nonumber\\
	\dot{m} &= &-(\gamma_m+i\Delta_m ) m - i g_{ma} a  - i g_{mb} m q + \Omega_l+ \xi e^{i\varphi} m^\dagger \nonumber \\ &+& \sqrt{2\gamma_m} m^\text{in}, \nonumber\\
	\dot{p} &=& -\omega_b q - \gamma_b p - g_{mb} m^\dagger m + \chi, \quad \dot{q} = \omega_b p, 
\end{eqnarray}
Where $a^{\mathrm{in}}$, $m^{\mathrm{in}}$, and $\chi$ are the zero-mean input noise operators of the microwave cavity, magnon, and phonon modes, respectively \cite{gardiner2004quantum}, in addition, their nonzero correlation functions are given by:
\begin{equation}
	\left\langle o^{in\dagger}(t)o^{in}(t') ; o^{in}(t)o^{in\dagger}(t') \right\rangle  = (N_o ; N_o + 1) \delta(t - t'),\, o=a,m.
\end{equation}
Moreover, $\chi$ represents the Brownian noise operator associated with the mechanical mode, which has a zero mean value, $\langle \chi \rangle = 0$. Under the Markov approximation, the corresponding correlation function for a mechanical oscillator with a high quality factor $Q \gg 1$ can be expressed as:
\begin{equation}
	 \left\langle \chi(t)\chi(t') + \chi(t')\chi(t)  \right\rangle/ 2 = \gamma_{b} (2N_{b} + 1) \delta(t - t').
\end{equation}
Here, the equilibrium mean thermal occupation numbers for the microwave cavity photons, magnons, and phonons of the YIG sphere are given by
$N_i = \left[\exp(\hbar \omega_i / k_B T) - 1 \right]^{-1}$ $(i = a, m, b)$,
where $k_B$ is the Boltzmann constant and $T$ is the environmental temperature.

 The quantum properties of the magnomechanical system are strongly influenced by the steady-state magnon and photon populations, which become much larger than unity under strong driving of both modes, i.e., $\left| \langle m \rangle \right| \gg 1$ and $\left| \langle a_s \rangle \right| \gg 1$ \cite{li2018magnon}. The nonlinear QLEs can then be expressed as the sum of the steady-state values and the corresponding quantum fluctuations, $O = O_s + \delta O$, where $O = a, m, q, p$. Accordingly, the steady-state mean values of the three-mode system are given by:
\begin{align}
m_s &= \frac{(\xi m_s^* e^{i\varphi} + \Omega_l)(\gamma_a+i\Delta_a)}{(\gamma_a+i\tilde {\Delta}_m )(\gamma_a+i\Delta_a) + g_{ma}^2}, \\
a_s &= \frac{-i g_{ma} m_s}{\gamma_a+i\Delta_a}, \\
q_s &= \frac{-i g_{mb} |m_s|^2}{\gamma_b+i\omega_b }, \quad p_s=0,
\end{align}
where $\tilde{\Delta}_m = \Delta_m + g_{mb}q_s$ represents the frequency shift induced by the magnomechanical interaction \cite{li2018magnon}. The steady-state magnon amplitude $m_s$ is explicitly given by:
 \begin{eqnarray}\label{Ems2}	
	m_{s}&=&  \langle m \rangle =\frac{\Lambda (\gamma_{a}+i{\Delta}_{a})+\xi(\gamma_{a}^2+{\Delta}_{a}^2)e^{i\varphi}}{ |\Lambda|^2 -\xi^2(\gamma_{a}^2+{\Delta}_{a}^2)}\Omega_l,
\end{eqnarray}
where  $\Lambda=(\gamma_a - i\Delta_a)(\gamma_m - i\tilde{\Delta}_m)+g_{ma}^2$.
Hence, under the condition $|\Delta_a|, |\tilde{\Delta}_m| \gg \gamma_a, \gamma_m$, the steady-state magnon amplitude $m_s$ in Eq.~(\ref{Ems2}) can be expressed as:
\begin{eqnarray}
m_{s} &=& \frac{\xi e^{i\varphi} + i\mu}{\mu^2 - \xi^2} , \Omega_l,
\end{eqnarray}
where $\mu = (g_{ma}^2 / \Delta_a) - \tilde{\Delta}_m$.

On the other hand, the linearized QLEs describing the system’s quantum fluctuations can be written in the following compact matrix form:
\begin{equation}
\dot{u}(t) = \Gamma u(t) + \mathcal{N}(t),
\end{equation}
where $u(t)^T = [\delta x_1(t), \delta y_1(t), \delta x_2(t), \delta y_2(t), \delta q, \delta p]$ is the quadrature vector.
The quadrature fluctuation operators are defined as:
$\delta x_1 = (\delta a + \delta a^\dagger)/\sqrt{2}$,
$\delta y_1 = (\delta a - \delta a^\dagger)/(i\sqrt{2})$,
$\delta x_2 = (\delta m + \delta m^\dagger)/\sqrt{2}$, and
$\delta y_2 = (\delta m - \delta m^\dagger)/(i\sqrt{2})$.
Moreover, $\mathcal{N}(t)^T = [\sqrt{2\gamma_a}x_1^{\mathrm{in}}(t), \sqrt{2\gamma_a}y_1^{\mathrm{in}}(t), \sqrt{2\gamma_m}x_2^{\mathrm{in}}(t), \sqrt{2\gamma_m}y_2^{\mathrm{in}}(t), 0, \chi]$ is the column vector of noise sources, and $A$ is the drift matrix given by:
\begin{equation}\label{E_A}
	A =\begin{pmatrix}
		-\gamma_{_a} & {\Delta}_a &0 & g_{ma} & 0 & 0  \\
		-{\Delta}_a & -\gamma_{a} &-g_{ma} & 0 & 0 & 0  \\
		0 & g_{ma} &-\gamma_{{\varphi_{+}}} & \tilde{\Delta}_{{\varphi_{+}}} & -G_{mb} & 0  \\
		-g_{ma} & 0 &-\tilde{\Delta}_{{\varphi{-}}} & -\gamma_{{\varphi_{-}}} & 0 & 0  \\
		0 & 0 &	0 & 0 & 0 & \omega_{b}  \\
		0 & 0 &	0 & G_{mb} & -\omega_{b} & -\gamma_{b}  \\
	\end{pmatrix},
\end{equation}
where $\tilde{\Delta}_{{\varphi_{\pm}}}=\tilde{\Delta}_{m}\pm {\Delta}_{\varphi}$ represents the effective detuning of the squeezed magnon field, incorporating the frequency shift induced by the magnomechanical interaction. Here, ${\Delta}_{{\varphi}}= \xi \sin\varphi$ and  $\tilde{\Delta}_{m}={\Delta}_{m}+ g_{mb} q_{s}$. In most cases, these frequency shifts are very small, satisfying $|\tilde{\Delta}_m-\Delta_m|\ll\Delta_m$ \cite{kong2019magnetically,lu2025nonreciprocal}. Moreover, $\gamma_{\varphi_{\pm}}=\gamma_m\pm \gamma_{\varphi}$ denotes the effective linewidth of the magnon mode, where  $\gamma_{\varphi}= \xi \cos\varphi$. The magnomechanical coupling strength, enhanced by coherent driving, is given by  $G_{mb}=i\sqrt{2}g_{mb}{m}_s$.

The steady-state quantum fluctuations form a Gaussian state, which can be completely characterized by the covariance matrix $\mathcal{V}$. This $6 \times 6$ real symmetric matrix has elements defined as
$\mathcal{V}_{ij} = \langle u_i(\infty)u_j(\infty) + u_j(\infty)u_i(\infty) \rangle / 2$ $(i, j = 1, 2, \ldots, 6)$,
where $u_{i(j)}(t)$ denotes the components of the fluctuation operator vector. In the steady-state limit $t \to \infty$, the covariance matrix $\mathcal{V}$ is determined by solving the Lyapunov equation \cite{la1976stability}:
\begin{equation}
	A\mathcal{V}+\mathcal{V} A^T+\mathcal{F}=0,
\end{equation}
where $\mathcal{F}$ is the diffusion matrix, given by
$\mathcal{F} = \mathrm{diag}[\gamma_a(2N_a + 1), \gamma_a(2N_a + 1), \gamma_m(2N_m + 1), \gamma_m(2N_m + 1), 0, \gamma_b(2N_b + 1)]$.


\section{Single-mode and total quantum coherence}\label{sec3}

Quantum coherence, a fundamental manifestation of the superposition principle, allows a system to exist in multiple states simultaneously and forms the basis of the unique features of quantum mechanics. It serves as an essential resource for quantum technologies and information processing. In the context of this work, we quantify the coherence of individual modes within the coupled system. In continuous variable systems, this coherence reflects correlations among quadrature fluctuations and plays a crucial role in the generation of entanglement.
To quantify this coherence, we consider a single-mode Gaussian state $\rho(\mathcal{V}_j, \vec{d}_j)$, where $\mathcal{V}_j$ ($j = a, m, b$) denotes the covariance matrix and the vector $\vec{d}_j = (d_{jx}, d_{jy})$ represents the mean values of the quadrature fluctuations, defined as $d_{jx} = (j_s + j_s^*) / \sqrt{2}$ and $d_{jy} = (j_s - j_s^*) / i\sqrt{2}$. The covariance matrix $\mathcal{V}_j$ for each subsystem is obtained by extracting the relevant block from the total covariance matrix $\mathcal{V}$. Accordingly, the covariance matrices corresponding to the photon, magnon, and phonon modes are given by:
\begin{equation}
	 \mathcal{V}_a=\begin{pmatrix}
		 \mathcal{V}_{11}& \mathcal{V}_{12}\\
		 \mathcal{V}_{21}& \mathcal{V}_{22}\\
	\end{pmatrix},
	 \mathcal{V}_m=\begin{pmatrix}
		 \mathcal{V}_{33}& \mathcal{V}_{34}\\
		 \mathcal{V}_{43}& \mathcal{V}_{44}\\
	\end{pmatrix}, 	 \mathcal{V}_b=\begin{pmatrix}
		 \mathcal{V}_{55}& \mathcal{V}_{56}\\
		 \mathcal{V}_{65}& \mathcal{V}_{66}\\ 
	\end{pmatrix}.
\end{equation}
The quantum coherence associated with the $j$-th mode can be quantified as \cite{xu2016quantifying}:
\begin{equation}
	C(\rho( \mathcal{V}_j,\vec d_{j}))=-h(v_j)+h(2\bar n_j+1),
\end{equation}
where $h(x)=(\frac{x+1}{2})\log_2(\frac{x+1}{2})-(\frac{x-1}{2})\log_2(\frac{x-1}{2})$, $\bar n_j=(\text{Tr}( \mathcal{V}_j)+d_{jx}^2+d_{jy}^2-2)/4$ and $v_j=\sqrt{\det( \mathcal{V}_j)}$ is the symplectic eigen value.

On the other hand, the total quantum coherence of the three-mode coupled system can be quantified as follows \cite{jin2021macroscopic}:
\begin{equation}
	C_T=C(\mathcal{V})=h(2\bar n_{a}+1)+h(2\bar n_{m}+1)+h(2\bar n_{b}+1)-   \sum_{k=1}^3h(w_k),
\end{equation}
where  $\bar n_{j}=\left(\text{Tr}( \mathcal{V}_j)+d_{jx}^2+d_{jy}^2-2 \right)/4 $ ($j=a,m,b$) is obtained from the covariance matrix $\mathcal{V}_j$ of the $j$th mode and its corresponding displacement vector $\vec{d}_j$. The set $\{w_k\}_{k=1}^3$ denotes the symplectic spectrum, which coincides with the ordinary eigenvalue spectrum of the matrix $|i \mathcal{S}  \mathcal{V}|$, where $ \mathcal{S}$ is the $6 \times 6$ symplectic form matrix:
\begin{equation}
    \mathcal{S} = \bigoplus_{k=1}^{3} \omega, \qquad \text{with} \qquad
\omega =
\begin{pmatrix}
0 & 1 \\
-1 & 0
\end{pmatrix}.
\end{equation}



\section{Numerical results and discussion}\label{sec4}

In this section, we present a numerical analysis of nonreciprocal quantum coherence arising from magnon-induced squeezing in the CMM system, using parameters obtained in current experiments. We focus on the behavior of the single-mode coherences ($C_a$), ($C_b$), ($C_m$), and the total coherence ($C_T$), then we examine how the driving power, magnon–photon coupling strengths, squeezing amplitude and phase, as well as environmental temperature influence nonreciprocity. This provides a clear picture of the tuning capability and control of quantum coherence in such systems. The parameters used are as follows \cite{lu2023magnon,li2018magnon,zhang2016cavity}:
$\omega_{a}/2\pi=\omega_{m}/2\pi=10$ GHz, $\omega_{m}/2\pi=10$ MHz, $\gamma_b/2\pi=100$ Hz, $\gamma_a=\gamma_m/5=2\pi\times 3$ MHz,  $g_{ma}/2\pi=4.8$ MHz, $g_{mb}/2\pi=0.1$ Hz, $P_l=0.3$ mW, $\Delta_a=\Tilde{\Delta}_m=\omega_b$ and $ T=10 $ K. Importantly, to guarantee a stable steady state in our numerical simulations, we apply the Routh–Hurwitz criterion \cite{dejesus1987routh}, requiring that all eigenvalues of the drift matrix have negative real parts.

	\begin{figure}[tbh!]
	\centerline{\includegraphics[width=13cm]{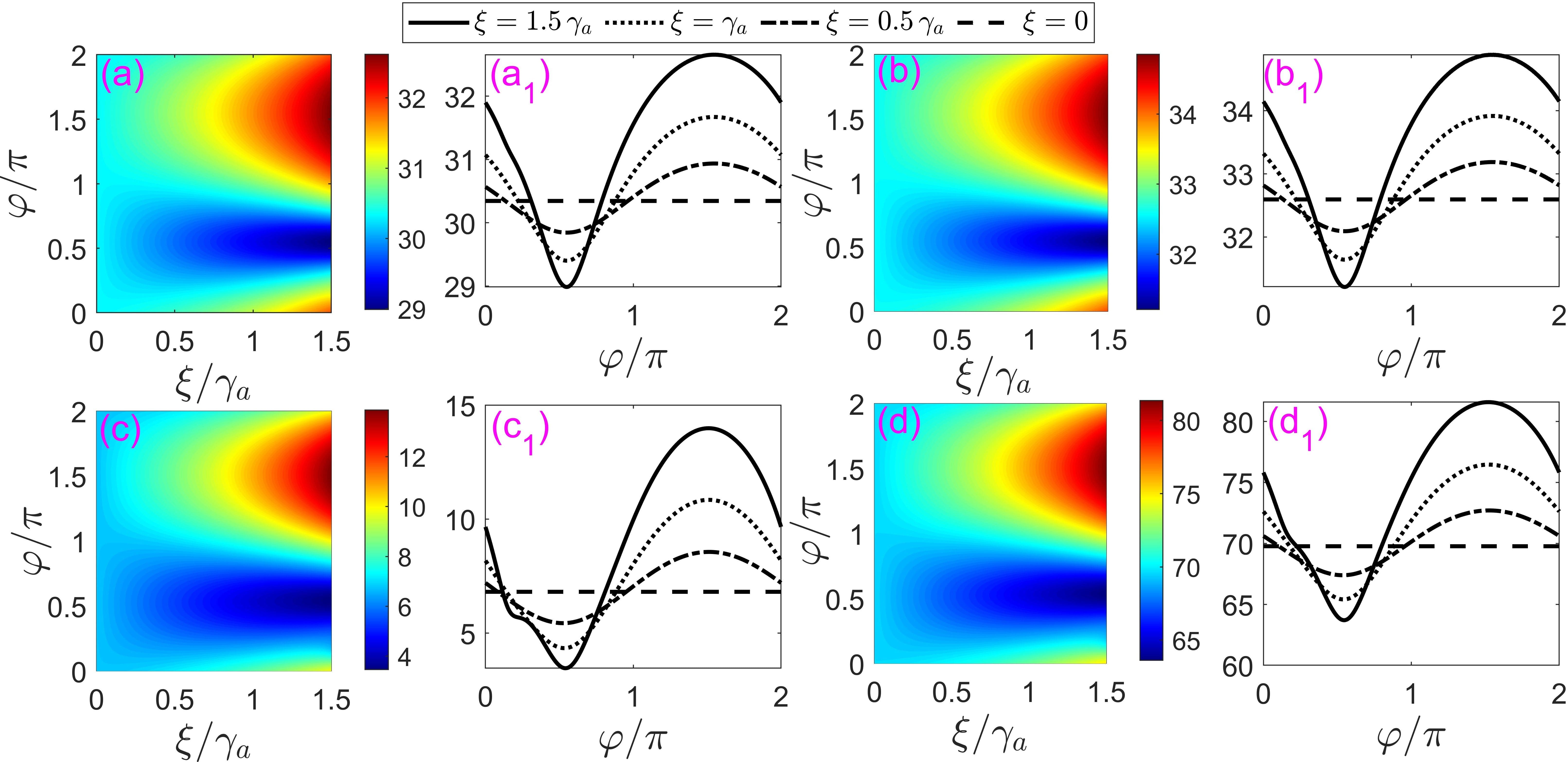}}
	\caption{\footnotesize{Density plots of quantum coherence: (a) $C_a$, (b) $C_m$, (c) $C_b$, and (d) $C_T$ as a function of the effective squeezing parameter $\xi$ and the phase $\varphi$. In addition, panels (a$_1$), (b$_1$), (c$_1$), and (d$_1$) show $C_a$, $C_m$, $C_b$, and $C_T$, respectively, as a function of the phase $\varphi$ for four values of the squeezing parameter: $\xi = 0$ (dashed line), $\xi = 0.5\gamma_a$ (dash–dotted line), $\xi = \gamma_a$ (dotted line), and $\xi = 1.5\gamma_a$ (solid line). The other parameters are given in the main text.
	}}
	\label{F2}
\end{figure}

In Fig. ~\ref{F2}(a)–(d), we plot the single-mode coherences ($C_a$, $C_m$, $C_b$) and the total coherence ($C_T$)  as a function of the normalized squeezing strength ($\xi/\gamma_a$) and the squeezing phase ($\varphi/\pi$). In all cases, the quantum coherence is phase-dependent. Indeed, the magnon squeezing term rotates the squeezed quadrature and simultaneously renormalizes the effective frequency and the magnon damping by phase-dependent shifts.  These shifts are given by $\Delta_\varphi = \xi\sin\varphi$ and $\gamma_\varphi = \xi\cos\varphi$. Together, these two control parameters reshape the steady-state covariance matrix and, hence, the resulting single-mode coherence.  Two phase angles are particularly important. At $\varphi = \pi/2$ and $\varphi = 3\pi/2$, the damping correction vanishes ($\gamma_\varphi = 0$), leaving only a pure frequency shift. For the bias point used here, a negative shift ($\varphi = 3\pi/2$) attracts the prepared magnon toward the operating resonance of the hybrid system. This maximizes the off-diagonal elements of the covariance matrix, creates global maxima of ($C_a$, $C_m$, $C_b$), and hence ($C_T$). Conversely, a positive shift ($\varphi = \pi/2$) moves the mode away, suppressing covariances and producing global minima. Away from these angles, $\gamma_\varphi \neq 0$ alters the effective linewidth. This introduces a damping-induced broadening, which causes a slight asymmetry in the coherence profiles near $\varphi = \pi$ and a gradual shift of the maximum-intensity ridge as $\varphi$ varies. Increasing $\xi$ enhances the phase contrast and raises the extremal coherence values to a broad optimum near $\xi \simeq 1.5\gamma_a$. Beyond this value, the enhancement is limited by a trade-off between improved coupling and increased effective linewidth or reduced stability margins. \\
Furthermore, Fig.~\ref{F2} (a$_1$)–(d$_1$) presents phase-dependent distributions that confirm this behavior for different values of $\xi$. For all modes, the coherence exhibits a minimum at $\varphi = \pi/2$ and a maximum at $\varphi = 3\pi/2$, with the peak-valley contrast increasing as $\xi$ increases. The absence of exact symmetry $\varphi \to \varphi + \pi$, where maxima do not reflect minima, indicates a direction-dependent redistribution of quantum fluctuations governed by the sign of $\Delta_\varphi$. The inversion of the phase $\varphi$ reverses the coherence flow between the three modes, thus revealing the nonreciprocal character of the system. In particular, compared to the stationary scenario ($\xi = 0$), quantum coherence is amplified when $\Delta_\varphi < 0$ and reduced when $\Delta_\varphi > 0$. This directional dependence indicates the emergence of nonreciprocal quantum coherence, which can only exist in the parameter regime where the system remains stable. All four panels show consistent behavior because the same phase-dependent covariance matrix elements determine the single-mode coherence of each mode. At the same time, $C_T$ reflects the global variation by integrating local and intermode correlation contributions.

	\begin{figure}[tbh!]
	\centerline{\includegraphics[width=13cm]{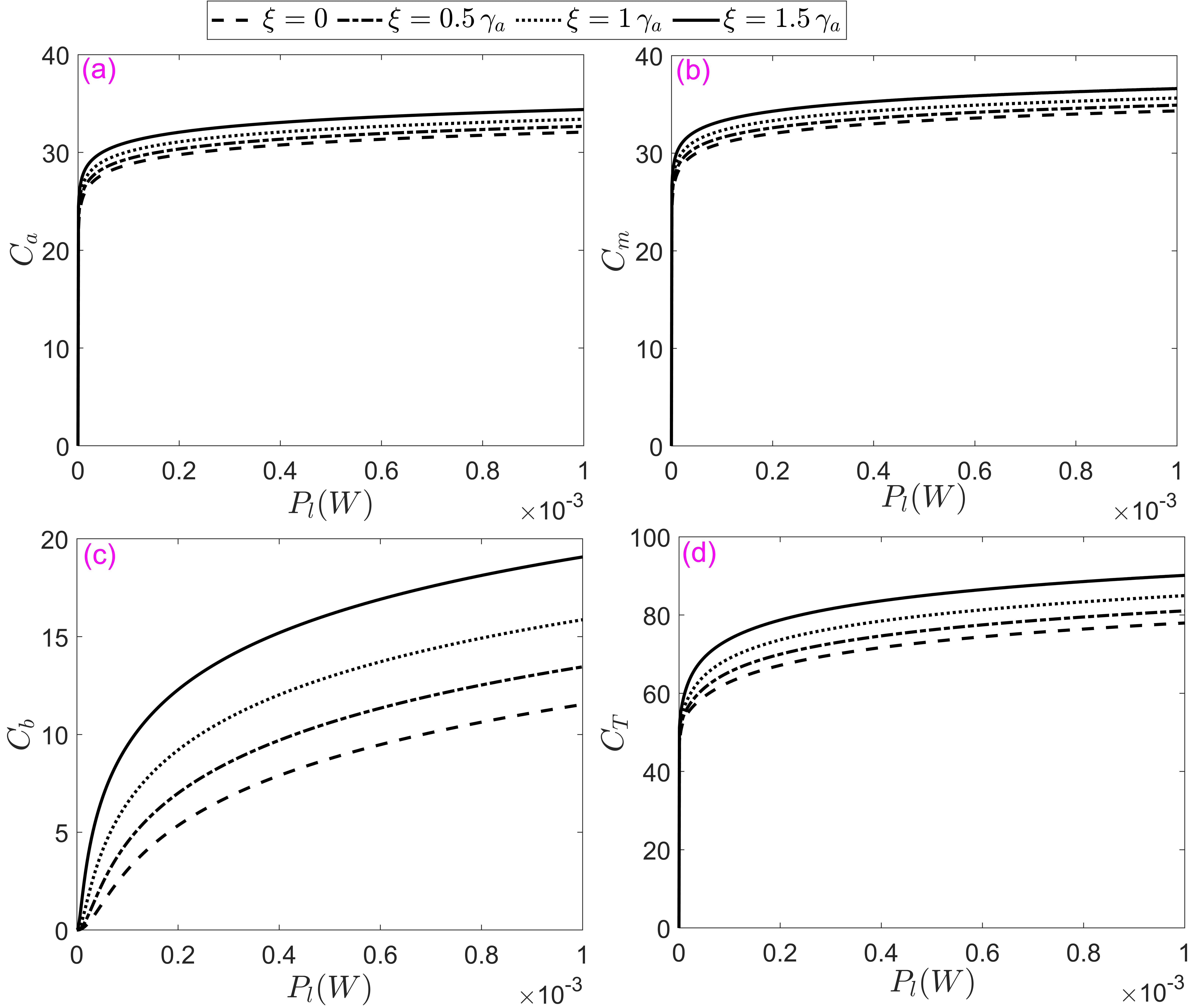}}
	\caption{\footnotesize{Plots of quantum coherence: (a) $C_a$, (b) $C_m$, (c) $C_b$, and (d) $C_T$ as a function of the driving power $P_l$ for four values of the squeezing parameter: $\xi = 0$ (dashed line), $\xi = 0.5\gamma_a$ (dash–dotted line), $\xi = \gamma_a$ (dotted line), and $\xi = 1.5\gamma_a$ (solid line). In all cases, the optimal phase is set to $\varphi = 3\pi/2$, and the other parameters are the same as those in Fig.~\ref{F2}.
	}}
	\label{F3}
\end{figure}

Figure~\ref{F3}(a)–(d) show the single-mode coherences ($C_a$, $C_m$, $C_b$) and the total coherence ($C_T$) as afunction of the driving power $P_l$ for four squeezing amplitudes $\xi$ (see legend). As $P_l$ increases, all coherences increase from zero to near-steady-state values, which is consistent with the linearized dynamics of the system. In particular, the steady-state amplitudes evolve as $|a_s|, |m_s| \propto \sqrt{P_l}$ and $|b_s| \propto P_l$, which increases the relevant covariance matrix elements and, hence, the Gaussian relative-entropy coherence of each mode \cite{lu2023magnon}. At higher powers, the resonances broaden and the system approaches its stability limits, preventing any improvement in coherence; consequently, the curves saturate and approach their steady-state values sublinearly.  In particular, among the three modes, the mechanical coherence $C_b$ increases most strongly because the drive enhances the magnon–phonon coupling. In the linearized regime, $G_{mb}=g_{mb} m_s/\sqrt{2}$; therefore, a larger $P_l$ increases $|m_s|$ and thus amplifies the transfer of quantum fluctuations into the mechanical quadratures. Meanwhile, $C_a$ and $C_m$ exhibit only moderate variation because, once $|a_s|$ and $|m_s|$ are balanced by $\gamma_a$ and $\gamma_m$, increasing $P_l$ mainly broadens the resonances instead of enhancing the correlations. Moreover, at fixed $P_l$, increasing the squeezing amplitude $\xi$ systematically improves quantum coherence due to the phase-dependent frequency shift generated by the squeezed-magnon term (Fig.~\ref{F2}). A larger $\xi$ brings the dressed magnon closer to resonance, thereby increasing $|m_s|$ and strengthening correlations between all modes. Consequently, $C_a$, $C_m$, $C_b$, and $C_T$ increase monotonically with $\xi$, with the strongest enhancement for $C_b$ (since $C_b \propto |m_s|^2$) and for $C_T$ which captures both single-mode and inter-mode contributions. Hence, Fig.~\ref{F3} shows that quantum coherence in this hybrid system can be deterministically controlled by two key parameters: the driving power $P_l$, which governs the overall coherence strength, and the squeezing amplitude $\xi$, which enables selective phase-dependent tuning of the system’s quantum correlations.

	\begin{figure}[tbh!]
	\centerline{\includegraphics[width=13cm]{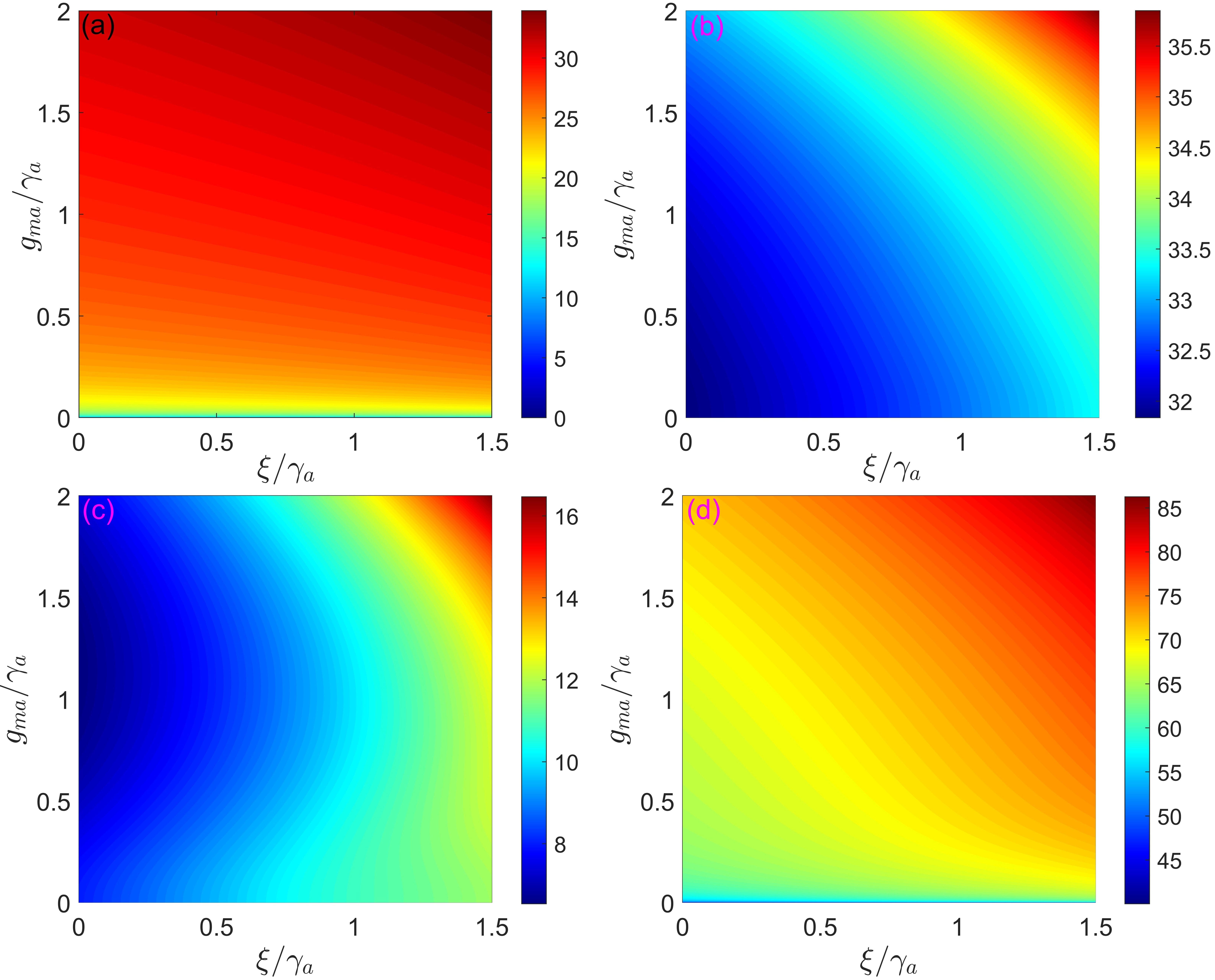}}
	\caption{\footnotesize{Density plots of quantum coherence: (a) $C_a$, (b) $C_m$, (c) $C_b$, and (d) $C_T$ as a function of the effective squeezing parameter $\xi$ and the coupling $g_{ma}$.  In all cases, the optimal phase is set to $\varphi = 3\pi/2$, and the other parameters are the same as those in Fig.~\ref{F2}.
	}}
	\label{F4}
\end{figure}

Figure~\ref{F4}(a)–(d) present the single-mode coherences ($C_a$, $C_m$, $C_b$) and the total coherence ($C_T$) as a function of the normalized squeezing amplitude $\xi/\gamma_a$ and the normalized photon–magnon coupling strength $g_{ma}/\gamma_a$, with the phase fixed at the value that optimizes coherence (i.e., $\varphi=3\pi/2$). The results reveal a monotonic increase in coherence with respect to the control parameters, attributable to two complementary physical processes. An increase in $g_{ma}$ enhances the hybridization between the cavity and magnon modes, thereby strengthening the cross-quadrature correlations and promoting the transfer of coherence to the mechanical subsystem. Simultaneously, a higher squeezing amplitude $\xi$ induces a phase-dependent frequency shift of the magnon, which enhances the steady-state magnon amplitude $|m_s|$ and consequently strengthens the effective magnon–phonon coupling $G_{mb} \propto |m_s|$. As a result, $C_a$ exhibits a dominant dependence on $g_{ma}$ with only weak sensitivity to $\xi$, while $C_m$ shows a pronounced dependence on both parameters, meaning their cooperative action. The mechanical coherence $C_b$ reaches its maximum value in the parameter range where $g_{ma}$ and $\xi$ are both high, confirming that substantial mechanical coherence only appears under conditions of strong photon–magnon coupling and significant squeezing. Total coherence $C_T$ exhibits a similar effect, as it integrates the collective influence of single-mode coherences as well as intermode correlations, reaching its maximum when $g_{ma}$ and $\xi$ jointly optimize the global quantum correlations while maintaining dynamical stability.

	\begin{figure}[tbh!]
	\centerline{\includegraphics[width=13cm]{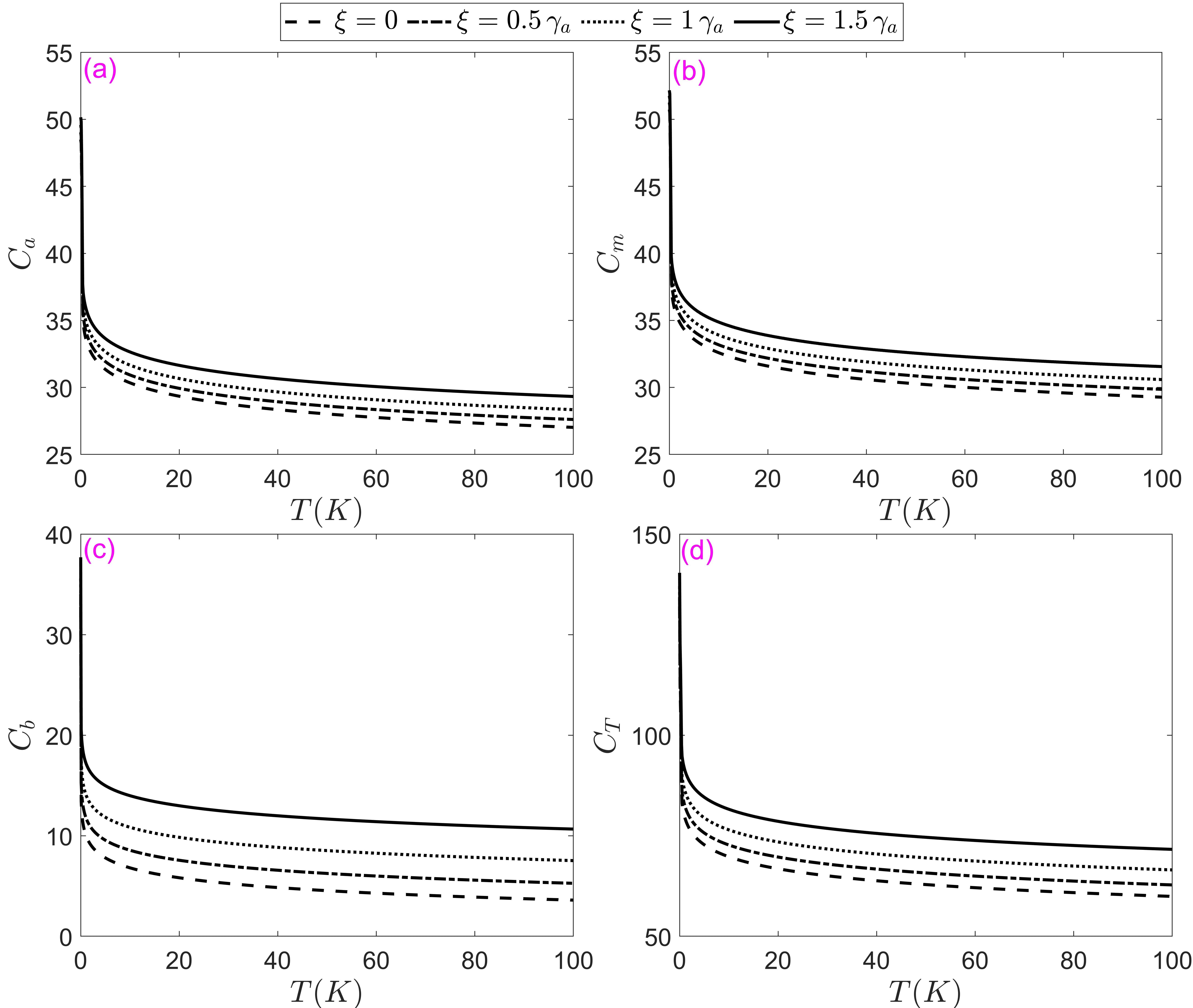}}
	\caption{\footnotesize{Plots of quantum coherence: (a) $C_a$, (b) $C_m$, (c) $C_b$, and (d) $C_T$ as a function of the environmental temperature $T$ for four values of the squeezing parameter: $\xi = 0$ (dashed line), $\xi = 0.5\gamma_a$ (dash–dotted line), $\xi = \gamma_a$ (dotted line), and $\xi = 1.5\gamma_a$ (solid line). In all cases, the optimal phase is set to $\varphi = 3\pi/2$, and the remaining parameters are the same as those in Fig.~\ref{F2}.
	}}
	\label{F5}
\end{figure}

Next, Fig.~\ref{F5}(a)-(d) present single-mode Gaussian coherences ($C_a$, $C_m$, $C_b$) and the total coherence ($C_T$) as a function of temperature $T$ for four squeezing amplitudes $\xi$ (see legend). We observe that all coherences decrease monotonically with $T$, exhibiting a steep decline at low temperatures followed by a gradual approach to an asymptotic value at higher $T$. This behavior arises from the thermal occupations $\bar{n}_o(T) = \big[\exp(\hbar\omega_o/k_BT) - 1\big]^{-1}$ ($o=a,m,b$), that enter into the diffusion matrix in the linearized dynamics. In particular, the mechanical contribution is proportional to $\gamma_b\big(2\bar{n}_b(T) + 1\big)$. It exerts the strongest influence on the global decoherence, since $\bar{n}_b(T) \approx k_BT/(\hbar\omega_b) \gg 1$ for MHz-frequency mechanical modes, even at moderate temperatures. Therefore, $C_b$ exhibits the highest thermal sensitivity, while $C_a$ and $C_m$ degrade more gradually with increasing $T$. Furthermore, increasing the squeezing amplitude $\xi$ enhances the overall coherence and improves its robustness to thermal noise at all temperatures. Indeed, when the compressed magnon term is applied to the optimal phase, the steady-state magnon amplitude $|m_s|$ increases, thus strengthening the coherent elements of the covariance matrix against incoherent thermal diffusion. This effect produces the observed vertical ordering $C(\xi = 1.5\,\gamma_a) > C(\xi = \gamma_a) > C(\xi = 0.5\,\gamma_a) > C(\xi = 0)$ in all panels, which is also confirmed by the previous results in Figs.~\ref{F2} and ~\ref{F3}. The total coherence $C_T$ remains consistently larger than the individual mode coherences, because intermode correlations persist even when each subsystem partially loses its coherence due to the thermal environment.

	\begin{figure}[tbh!]
	\centerline{\includegraphics[width=13cm]{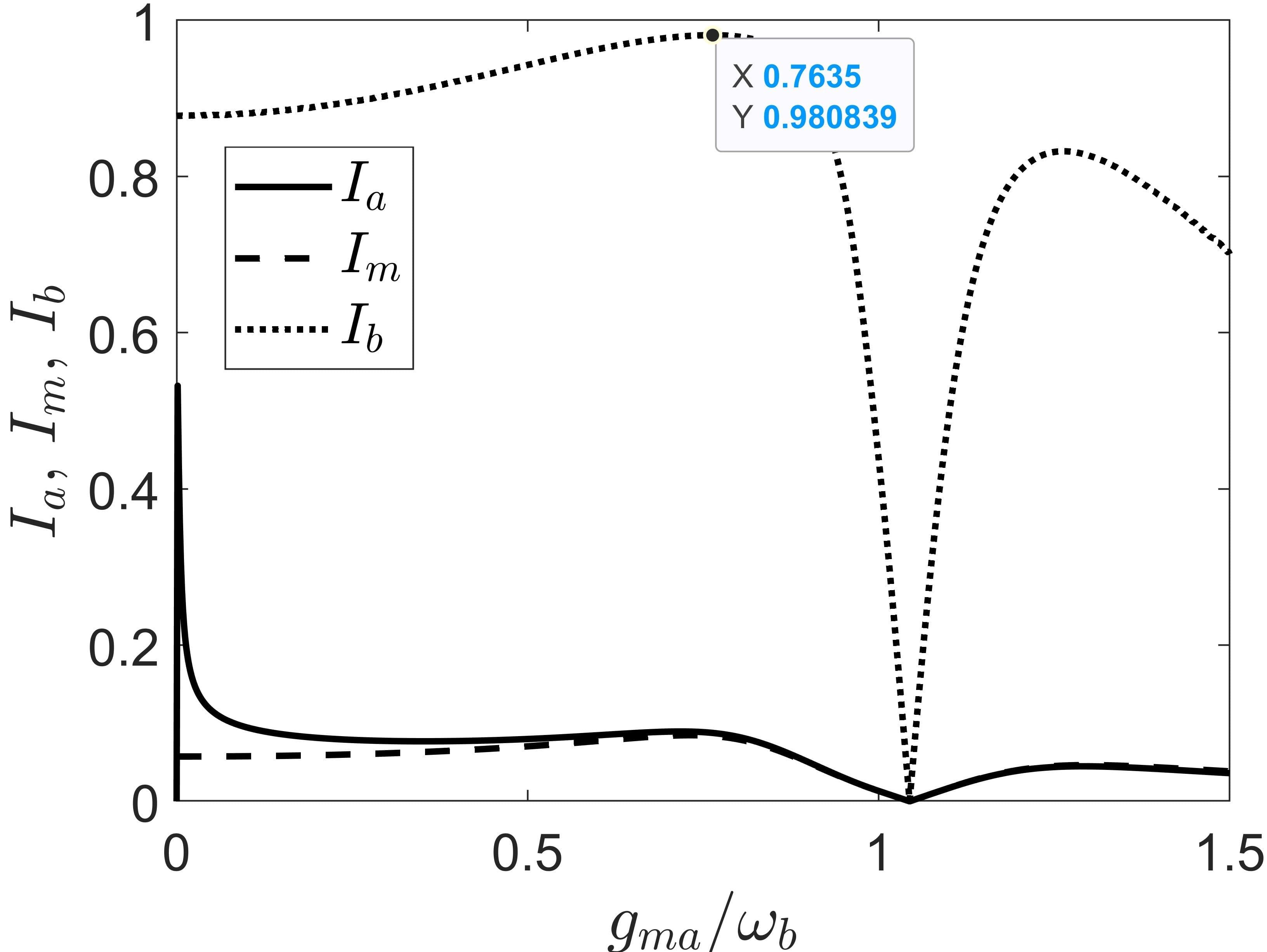}}
	\caption{\footnotesize{Plots of bidirectional contrast ratio: $I_a$ (solid line), $I_m$ (dashed line), and $I_b$ (dotted line) as a function of $g_{ma}$. In all cases, the squeezing parameter is set to $\xi = 1.1\gamma_a$, the driving power to $P_l = 1\times10^{-4}$ mW, and the remaining parameters are the same as those in Fig.~\ref{F2}.
	}}
	\label{F6}
\end{figure}

To precisely quantify nonreciprocal quantum coherence, we introduce the bidirectional contrast ratio ($I_i $), which serves as a quantitative indicator of nonreciprocity, defined as follows \cite{imara2025squeezed,lu2025nonreciprocal}:
            \begin{equation}
                I_i=\frac{\big|C_i{(\Delta_\varphi>0)}-C_i{(\Delta_\varphi<0)}\big|}{C_i{(\Delta_\varphi>0)}+C_i{(\Delta_\varphi<0)}},\quad (i=a,m,b),
            \end{equation}
where $C_i$ are the single‑mode coherences computed for two opposing squeezing phases that flip the sign of the magnon’s phase‑induced shift.%

Figure~\ref{F6} shows the bidirectional contrast ratios $I_a$, $I_m$, and $I_b$ versus the normalized photon–magnon coupling $g_{ma}/\omega_b$. The curves are computed for two squeezing phases of opposite signs, which reverse the magnon’s phase-induced frequency shift. This definition follows the nonreciprocity metric introduced for coherence in Ref. \cite{jia2025nonreciprocal}, where the Barnett-induced frequency shift is replaced here by a phase-controlled shift generated by magnon squeezing. When $g_{ma}=0$, $I_a$ remains zero, indicating 
that there is no nonreciprocal quantum coherence and that forward and backward propagation are totally symmetric. When $g_{ma}$ increases, $I_a$ increases sharply and then gradually decreases, resulting from the transition of the photon mode from a decoupled state to a weakly coupled regime with the magnon mode. Furthermore, when $g_{ma}\approx\omega_b$,  we note a complete suppression of nonreciprocal quantum coherence in all modes, marking a reciprocity point. This behavior can be understood through the steady-state analysis of the system. Under the condition $|\tilde{\Delta}_m|,|\Delta_a|\gg\gamma_a,\gamma_m$, the steady-state magnon amplitude $m_s$ can be simplified and expressed as: $m_s\approx\Omega_l (\xi e^{i\varphi}+i\mu)/(\mu^2-\xi^2)$ with $\mu=(g_{ma}^2/\Delta_a)-\tilde{\Delta}_m$. When the squeezing is deactivated ($\xi=0$), that is, when the YIG sphere is no longer squeezed, the system ceases to exhibit steady-state nonreciprocity. The minimal nonreciprocity appears near the hybridization resonance condition $g_{ma}^2\simeq\Delta_a\tilde{\Delta}_m$, corresponding to $g_{ma}\simeq\omega_b$. As $g_{ma}/\omega_b$ increases from $0$ to $1$, the mechanical (phonon) contrast remains close to $1$, reaching a maximum value of approximately $0.9808$. These results highlight the central role of the photon–magnon interaction in governing nonreciprocal quantum coherence and demonstrate that magnon squeezing enables pronounced phonon nonreciprocity on a microscale platform, offering a promising path toward the on-chip integration of nonreciprocal phononic elements.


	\section{Conclusion}\label{sec5}

In this work, we have theoretically analyzed quantum coherence in a cavity magnomechanical system incorporating squeezed magnons. The results show that magnon squeezing introduces a modulation, dependent on the phase, effective frequency, and damping of the magnon mode, allowing precise control of the coherence properties of the hybrid system. The driving power and the photon–magnon coupling strength are identified as key parameters governing the overall coherence level and its distribution between the cavity, magnonic, and mechanical subsystems. Increasing the squeezing amplitude enhances the steady-state magnon amplitude and strengthens coherent energy exchange between the coupled modes, thus improving total coherence. Moreover, the results show that squeezed magnons significantly increase the robustness of coherence against thermal noise and induce direction-dependent (nonreciprocal) behavior in the coherence flow between subsystems. The results establish magnon squeezing as an efficient and tunable mechanism for controlling nonreciprocal and thermally resistant quantum coherence in cavity magnomechanical platforms.



\section*{Declarations} 
\textbf{Competing Interests:} The authors declare no competing interests

\bibliography{sn-bibliography}

\end{document}